\documentclass[aip,jmp,amsmath,showkeys,showpacs,preprint]{revtex4-1}

\usepackage{graphicx}
\usepackage{bm}

\begin{document}

\title{Wavefunction collapse through backaction of counting weakly interacting photons}

\author{L. E. Harrell}
\email{lee.harrell@usma.edu}
\thanks{The views expressed herein are those of the author and do not reflect
the position of the Department of the Army or the Department of Defense.}
\affiliation{Department of Physics and Nuclear Engineering, United States
Military Academy,
West Point, NY, USA}

\date{\today}

%%%%%%%%%%%%%%%%%%%%%%%%%%%%%%%%%%%%%%%%%%%%%%%%%%%%%%%%%%%%%%%%%
\begin{abstract}
We apply the formalism of quantum measurement theory to the idealized measurement of the position of a particle with an optical interferometer, finding that the backaction of counting entangled photons systematically collapses the particle's wavefunction toward a narrow Gaussian wavepacket at the location $x_\mathrm{est}$ determined by the measurement without appeal to environmental decoherence or other spontaneous collapse mechanism.
Further, the variance in the particle's position, as calculated from the post-measurement wavefunction agrees precisely with shot-noise limited uncertainty of the measured $x_\mathrm{est}$.
Both the identification of the absolute square of the particle's initial wavefunction as the probability density for $x_\mathrm{est}$ and the de Broglie hypothesis emerge as consequences of interpreting the intensity of the optical field as proportional to the probability of detecting a photon.
Linear momentum information that is encoded in the particle's initial wavefunction survives the measurement, and the pre-measurement expectation values are preserved in the ensemble average.
\end{abstract}



\maketitle

%%%%%%%%%%%%%%%%%%%%%%%%%%%%%%%%%%%%%%%%%%%%%%%%%%%%%%%%%%%%%%%%%
\section{\label{sec:Introduction}Introduction}
There is a problem at the foundation of quantum mechanics.
Among its postulates are two that make seemingly contradictory assertions about the time evolution of the physical state of a system.
According to one postulate, the state evolves according the deterministic Schr{\"o}dinger equation.
The other postulate in question asserts that the state changes randomly whenever a measurement occurs, collapsing to a new state that is consistent with the random outcome of the measurement.\cite{Luders2006sep, Ruza2010jan, Allahverdyan2013apr}
%Attempts to treat both postulates as fundamental challenge the concept of a physical world with an observer-independent existence that forms the basis of our common experiences and in which observers and measuring devices obey the same physical laws as the systems they study---a state of affairs in quantum mechanics commonly known as the measurement problem.
Attempts to treat both postulates as fundamental challenge the concept of a physical world with an observer-independent existence that forms the basis of our common experiences and in which observers and measuring devices obey the same physical laws as the systems they study---a state of affairs in quantum mechanics commonly known as the measurement problem.\cite{Bassi03, Schlosshauer04, Weinberg2012jun, Bassi13}

In practice, the application of quantum mechanics requires a choice of which time-evolution postulate to use.
In spite of the fact that the theory is silent on how to determine when a measurement has or has not occurred, physicists reliably make the correct choice, leading many to conclude that
either the problem does not exist or its serious consideration is not urgent.\cite{mermin89}
Others have made concerted efforts to resolve the contradiction by reformulating or reinterpreting the theory.
Current lines of development include spontaneous wave function collapse models,\cite{Nimmrichter2014jul, Diosi2015feb} the many worlds, consistent histories\cite{Hohenberg2010oct} and quantum-Bayesian (QBist) interpretations,\cite{Fuchs13} and environmental decoherence theory\cite{Zurek2003may, Pernice2011dec}.

The archetypal illustration of the measurement problem is the collapse of a particle's wavefunction during an optical measurement of its position.
Wavefunction collapse also occurs in the collision of a particle with a detector, but in this case the unitary evolution of the particle's state under its free-particle Hamiltonian must come to a messy end on reaching the detector.
Accounting for the macroscopic detector's internal degrees of freedom is intractable, so it is not possible to compare evolution of the particle state under the measurement postulate and evolution of the particle-detector system under the Schr{\"o}dinger equation.
On the other hand, the light field in the optical measurement might be described by inclusion of an interaction term in the Hamiltonian, and the state could continue to evolve under the Schr{\"o}dinger equation.
Alternatively, the particle could scatter light and mysteriously appear at a specific place and time with a collapsed wave function characterized by a small uncertainty in position and a large uncertainty in linear momentum in accordance with the measurement postulate, which gives us no insight into the mechanism by which the wavefunction evolved to this new state or how its position came to be correlated to the observed distribution of the scattered light.\cite{Parisio2011dec}
Quantum mechanics does not seem to favor one alternative over the other.

The present work does not attempt to resolve the measurement problem.
Rather, it presents an idealized model to demonstrate that the distinction between the mysterious in-flight wavefunction collapse and the collapse that occurs when a particle strikes a detector is not fundamental.
In the model, the mechanism of wavefunction collapse is the creation of quantum entanglement between the particle and photons and its subsequent destruction when the photons are detected.
Given the data record from the photon detectors, the model makes a quantitative prediction for the post-measurement particle wavefunction that approaches the predictions of the measurement postulate in the limit of a precise position measurement.

Remarkably, the model is developed from optical devices and quantum concepts---interferometers, photon counting, two-particle entanglement---at
the level of many modern physics courses.
As such, it could be incorporated into the modern physics narrative to provide a conceptual bridge between the intensity of the optical field as a measure of the photon detection probability and the absolute square of the nonrelativistic wavefunction as the probability density for the position of a particle.
The model also motivates the de Broglie hypothesis for the wavelength of a nonrelativistic particle.

The remainder of this paper is organized as follows. Sec.\ \ref{sec:Position} introduces a specific interferometer configuration, provides an analysis of its use to measure the position of a classical particle by photon counting, and discusses the fundamental limits placed on such a measurement by photon shot noise.
In Sec.\ \ref{sec:Collapse} we develop a description of the same position measurement in terms of quantum measurement theory and repeated weak measurements with an emphasis on the quantum trajectory of the particle's state toward a collapsed eigenstate of the position operator.
The statistics of an ensemble of such measurements are considered in Sec.\ \ref{sec:Averages} and Sec.\ \ref{sec:Momentum}.   Section \ref{sec:WeakValues} discusses the relationship to observations of weak values, followed by concluding remarks in Sec.\ \ref{sec:Conclusion}.

%%%%%%%%%%%%%%%%%%%%%%%%%%%%%%%%%%%%%%%%%%%%%%%%%%%%%%%%%%%%%%%%%
\section{\label{sec:Position}Interferometer-based position estimate and position uncertainty}

\begin{figure}
\includegraphics{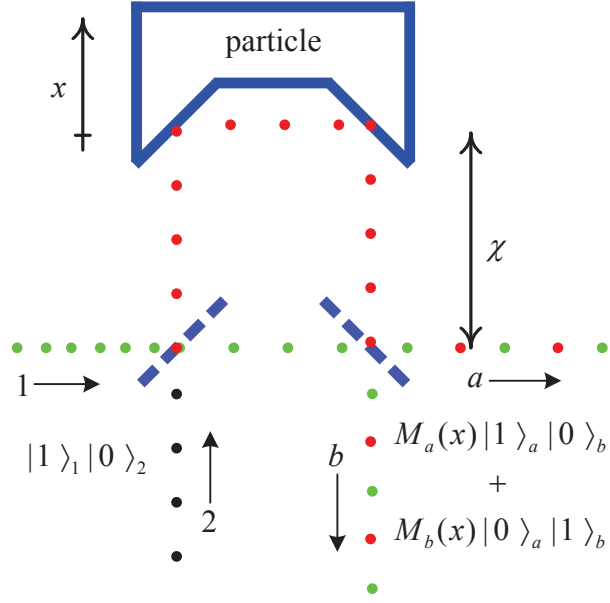}
\caption{\label{fig:interferometer} (Color online)
An ideal interferometer splits an incoming ray, allows the split rays to propagate along two distinct paths and then recombines the rays to produce interference effects.
The interferometer considered in this paper consists of a monochromatic photon source, two symmetric beam splitters, two mirrors, and two photon-counting detectors.
The two mirrors are rigidly incorporated into the particle whose position we wish to measure.  Photons are introduced at port 1, and the vacuum feeds port 2.
Output photons are detected at ports $a$ and $b$.
In a properly tuned interferometer, estimation of the probabilities with which photons arrive at each detector by passing a large number $N$ of photons through the device constitutes a measurement of the distance $\chi$ or, equivalently, the position $x$.
}
\end{figure}

Any interferometer-based measurement correlates the value of the observable being measured with the excess phase of light that is transmitted through one of the interferometer's optical paths.
Small interferometers are routinely used as sensors of position and other observables in experiments.\cite{Fleming13}
The essence of measuring the position of a particle is establishing the distance along a particular direction between the particle and a reference point.
For the interferometer-based position measurement described in Fig.\ \ref{fig:interferometer}, the location of a beam splitter serves as the reference point.

For this Mach-Zehnder configuration, the position of the particle relative to the beam splitter and the length of the interferometer's upper arm are then identical.
The length of this arm determines the excess phase in the upper optical path of the interferometer, thereby allowing the measurement of the particle's position through interference with light in the lower reference path.
Operationally, a position measurement consists of sequentially passing $N$ photons of wavenumber $k$ through the interferometer and recording their arrival at the detectors.
We ignore the evolution of the particle's state under its internal Hamiltonian for the short duration of the measurement.

The effect of transport through the interferometer on the photon is determined through a straightforward application of physical optics.
Adopting Loudon's notation and phase convention for symmetric beam splitters,\cite{Loudon03} the state of a single photon introduced to the input of the interferometer by the photon source at port 1 and the vacuum at port 2,
\begin{equation}
|\mathrm{in}\rangle = |\mathrm{1}\rangle_1|\mathrm{0}\rangle_2,\label{eq:interferometerin}
\end{equation}
is written in the output basis as
\begin{equation}
|\mathrm{out}\rangle =
\frac{{e^{i2k\chi} + 1}}{{2}}|\mathrm{1}\rangle_a|\mathrm{0}\rangle_b +
\frac{{e^{i2k\chi} - 1}}{{2i}}|\mathrm{0}\rangle_a|\mathrm{1}\rangle_b.
\label{eq:interferometerout}
\end{equation}
The amplitudes for detecting a photon in each output port vary with $\chi$, and the corresponding probabilities oscillate sinusoidally.
The instrument is most sensitive to the position of the particle (i.e. the slope of the photon detection probability as a function of $\chi$ is maximum) when $\chi \approx {\pi}(n + 1/4)/k$, where $n$ is an integer.
Taking one of these sensitive points as the origin of the $x$-axis, we make the substitution $\chi = x + {\pi}(n + 1/4)/k$ in Eq.\ (\ref{eq:interferometerout}).
The state of the photon in the output basis is then written as
\begin{equation}
|\mathrm{out}\rangle = M_a(x)|\mathrm{1}\rangle_a|\mathrm{0}\rangle_b +
M_b(x)|\mathrm{0}\rangle_a|\mathrm{1}\rangle_b,
\label{eq:TunedInterferometerTransfer}
\end{equation}
where
\begin{equation}
\begin{split}
M_a(x) &=  \frac{i\exp(i2kx) + 1}{2}\\
M_b(x) &=  \frac{i\exp(i2kx) - 1}{2i}.
\end{split}\label{eq:MeasurementOperators}
\end{equation}
Tuning of the interferometer is achieved by adjusting $k$ or the spacing between the particle and the beam splitters so that the nominal position of the particle is near $x=0$.

The probability of detecting a photon at each output port is the absolute square of the amplitude for the photon to be in the corresponding state:
\begin{equation}
\begin{split}
P_a &= \left|M_a(x)\right|^2 = \frac{1}{2}\left(1 - \sin{2 k x}\right), \\
P_b &= \left|M_b(x)\right|^2 = \frac{1}{2}\left(1 + \sin{2 k x}\right). \\
\end{split}\label{eq:channelprobability}
\end{equation}
In the limit of small $kx$ (i.e., in the case of a tuned interferometer where the particle is localized in a region  much smaller than one photon wavelength from its nominal position), the difference between the probability for a photon to be detected at port $b$ and port $a$ is related to $x$ by
\begin{equation}
x = \frac{{P_b  - P_a }}{{2k}}. \label{eq:InterferometerGain}
\end{equation}
The position of the particle is estimated by passing $N$ photons through the interferometer and counting the number of photons $n_a$ and $n_b$ detected at each port.
Then,
\begin{equation}
x_{\mathrm{est}} = \frac{n_b - n_a}{2kN}.
\label{eq:xest}
\end{equation}

For finite $N$, $n_a/N$ and $n_b/N$ are only approximations of the probabilities $P_a$ and $P_b$, so the precision of the position estimate is limited by the shot noise of the photons.
From the observation that tuning implies $|n_a - n_b| << N$, the variance of $x_{\rm{est}}$ is
\begin{equation}
\sigma_{x_{\mathrm{est}}}^2 = \frac{1}{4 N k^2}.
\label{eq:VarianceXEstimate}
\end{equation}

Even though $x$ is taken to be a continuous variable, the measurement can only return one of $N+1$ discrete values for $x_\mathrm{est}$. The smallest possible increment in $x_\mathrm{est}$ is
\begin{equation}
\delta x_{\mathrm{est}} = \frac{1}{k N}.
\label{eq:binxest}
\end{equation}
In other words, the result of the measurement is that $x$ was found to be in the range $(x_\mathrm{est} - \delta x_{\mathrm{est}}/2, x_\mathrm{est}+\delta x_{\mathrm{est}}/2)$.
While it may be tempting to make a direct comparison between $\sigma_{x\rm{ est}}$ and $\delta x$, it is clear from their different scaling with $N$ that they must have different physical interpretations.

%%%%%%%%%%%%%%%%%%%%%%%%%%%%%%%%%%%%%%%%%%%%%%%%%%%%%%%%%%%%%%%%%
\section{\label{sec:Collapse}Collapse of the wavefunction}

From the experimenter's point of view, interpretation of the interferometer data is independent of whether a classical or quantum particle is observed.
The question of interest in this section is whether orthodox quantum mechanics,\cite{vonN32} at least in the limit of $N\rightarrow\infty$, provides a mechanism by which the particle' quantum state collapses systematically toward an eigenstate of position at the location $x_\mathrm{est}$ with position uncertainty $\sigma_{x\rm{ est}}$ and probability $\left|\psi_0(x_{\mathrm{est}})\right|^2\delta x$ as a consequence solely of photon detection at the two output ports.
To answer this question, we must determine how the composite quantum state of the particle-photon system evolves between the introduction of each photon and its subsequent detection.
We do so in the context of a generalized quantum measurement.\cite{Nielsen00, Wiseman09, Clerk2010apr, Jacobs14}

In a generalized quantum measurement, the system being measured (the target), initially in the quantum state $|\psi\rangle$, couples via unitary evolution $U$ under a transient interaction Hamiltonian with an ancillary system (the probe), which has been prepared independently in the quantum state $|i\rangle$.  Prior to the interaction, the uncorrelated state of the composite target-probe system is
\begin{equation}
|\Psi\rangle=|\psi\rangle\otimes|i\rangle.
\label{eq:generalinital}
\end{equation}
Following the interaction, the composite system is left in the entangled state
\begin{align}
|\widetilde{\Psi}\rangle &= U|\psi\rangle\otimes|i\rangle\nonumber\\
&= \sum_n M_n |\psi\rangle\otimes|n\rangle,
\label{eq:generalintermediate}
\end{align}
where the probe state has been expressed in terms of the measurement basis $\{|n\rangle\}$, and the measurement operators ${M_n}$, which are determined by $U$, $|i\rangle$ and $\{|n\rangle\}$, act on the state space of the target.
Subsequently, a projective measurement in the basis $\{|n\rangle\}$ is made on the probe, leaving the target in the state
\begin{equation}
|\widetilde{\psi}\rangle = \frac{M_n |\psi\rangle}{\sqrt{\langle\psi|M_n^\dagger M_n|\psi\rangle}}
\label{eq:generalfinaltarget}
\end{equation}
with probability
\begin{equation}
P_n=\langle\psi|M_n^\dagger M_n|\psi\rangle.
\label{eq:generalprob}
\end{equation}
The general theory does not indicate what property of the target has been determined by the projective measurement on the probe or the result of the backaction of the probe measurement on the target.  Both of these determinations must be inferred from the detailed physics of the probe and the probe-target interaction.

Of particular interest are discrete, or even continuous, iterative sequences of weak measurements applied repeatedly to a single quantum system.
It is well-known that strong or projective measurements can be decomposed, at least formally, into a sequence of
weak measurements that generate a trajectory in Hilbert space from the initial system state to its post measurement state.\cite{Oreshkov05, Varbanov2007sep}
The study of such sequences of weak measurement underlies many lines of research, including quantum stochastic processes,\cite{Gisin1984may, Gisin1989, Belavkin1992jun, Brun2002unk} quantum information theory,\cite{Cheong2012oct} weak value amplification,\cite{Aharonov1988apr, Ritchie1991mar, Dixon2009apr, Starling2009oct, Nishizawa2012jun, Rozema2012sep, Lund2010sep} and quantum feedback and control.\cite{Wiseman09, DAriano1997jul}

We interpret the interferometric position measurement of Fig.\ \ref{fig:interferometer} as an iterative sequence of $N$ weak generalized quantum measurements that proceeds one photon at a time and evolves the wavefunction of the particle from $\psi_0(x)= \langle x|\psi_0\rangle$ to $\psi_N(x)= \langle x|\psi_N\rangle$.
For each iteration, the target is the particle in quantum state $|\psi_j\rangle$, while the probe is the optical system, which is initialized in the state $|i\rangle = |\mathrm{1}\rangle_1|\mathrm{0}\rangle_2$.
The measurement basis for the probe is
\begin{equation}
\left\{|a\rangle = |\mathrm{1}\rangle_a|\mathrm{0}\rangle_b, |b\rangle = |\mathrm{0}\rangle_a|\mathrm{1}\rangle_b\right\},
\label{eq:measurementbasis}
\end{equation}
and the projection of the probe state onto this basis is accomplished and recorded by the photon detectors at the output ports of the interferometer.

Prior to the $j^{\mathrm{th}}$ measurement, the state of the probe-target system is then
\begin{equation}
|\Psi\rangle=|\psi_{j-1}\rangle\otimes|\mathrm{1}\rangle_1|\mathrm{0}\rangle_2.
\label{eq:jthinital}
\end{equation}
The entangled state of the probe-target system after the interaction of the photon with the particle, but prior to the projection of the probe onto the measurement basis, follows from the same physical-optics reasoning that led to Eqs.\ (\ref{eq:TunedInterferometerTransfer}) and (\ref{eq:MeasurementOperators}), with the caveat that the optical path length of the upper arm must be interpreted as a function of the particle's position operator, implying
\begin{align}
|\widetilde{\Psi}\rangle &= M_a(x)|\psi_{j-1}\rangle\otimes|a\rangle + M_b(x)|\psi_{j-1}\rangle\otimes|b\rangle\nonumber\\
&= c_{aj}\frac{M_a(x)|\psi_{j-1}\rangle}{c_{aj}}\otimes|a\rangle+c_{bj}\frac{M_b(x)|\psi_{j-1}\rangle}{c_{bj}}\otimes|b\rangle,
\label{eq:jthintermediate}
\end{align}
where the normalization factors are
\begin{align}
c_{aj}  &= \sqrt{\langle\psi_{j-1}|M_a^{\dagger}(x)M_a(x)|\psi_{j-1}\rangle}\nonumber\\
c_{bj}  &= \sqrt{\langle\psi_{j-1}|M_b^{\dagger}(x)M_b(x)|\psi_{j-1}\rangle}.
\label{eq:ChannelProbabilities}
\end{align}
Comparison of Eq.\ (\ref{eq:jthintermediate}) with Eq.\ (\ref{eq:generalintermediate}) shows that $M_a(x)$ and $M_b(x)$, which are given by Eq.\ (\ref{eq:MeasurementOperators}) but are now understood to be functions of the position operator, can be identified as the measurement operators in each iteration.
In what follows, we will show a remarkable consistency of the interpretation of the data as indicating the position of the particle, the form of the particles's post-measurement wave function, and the probability distribution of final states calculated from the particle's pre-measurement wavefunction, giving a powerful justification for this identification.

Subsequently, the $j^{\mathrm{th}}$ photon is detected at either port
$a$ or port $b$.  The photon is destroyed and the particle is left in the
state
\begin{subequations}
\label{eq:PerturbedWaveFunction}
\begin{equation}
|\psi_{j}\rangle = \frac{M_a(x)|\psi_{j-1}\rangle}{c_{aj}}
\end{equation}
or
\begin{equation}
|\psi_{j}\rangle = \frac{M_b(x)|\psi_{j-1}\rangle}{c_{bj}}
\end{equation}
\end{subequations}
with probability $P_{aj}=|c_{aj}|^2$ or $P_{bj}=|c_{bj}|^2$ respectively.

The data record for a single measurement consists of an ordered sequence of photon counts from the two detectors of the form
\begin{equation}
(o_1,o_2,o_3, \cdots )=(a,a,b,a,b,a,b,b,b, \cdots ).
\end{equation}
For a particle that starts in the state $|\psi_0\rangle$, the normalized state
immediately after the measurement is
\begin{equation}
|\psi_N\rangle = \frac{O_N \cdots O_2O_{1}|\psi _0\rangle}
{\sqrt{\langle\psi_0|O_{1}^\dagger O_{2}^\dagger \cdots O_N^\dagger O_N \cdots O_{2}O_{1}|\psi _0\rangle}},
\label{eq:FinalWaveFunction1}
\end{equation}
where the $O_j \in \{M_a(x), M_b(x)\}$ and are selected according to the data
record.  As functions of position, the measurement operators commute, so Eq.\
(\ref{eq:FinalWaveFunction1}) immediately simplifies to
\begin{equation}
|\psi_N\rangle = \frac{{M_a(x)^{n_a } M_b(x)^{n_b }|\psi_0\rangle}}
{\sqrt{\langle\psi_0|(M_a^\dagger(x)M_a(x))^{n_a}(M_b^\dagger(x)M_b(x))^{n_b}|\psi_0\rangle}}
\label{eq:FinalWaveFunction2}
\end{equation}
for \emph{any} data sequence with $n_a$ photons detected at port $a$ and
$n_b=N-n_a$ photons detected at port $b$. Further, using Eq.\ (\ref{eq:xest}) to
write $n_b - n_a$ in terms of $x_{\mathrm{est}}$ and Eq.\
(\ref{eq:MeasurementOperators}) for $M_a(x)$ and $M_b(x)$ we have
\begin{equation}
\begin{split}
M_a(x)&^{n_a } M_b(x)^{n_b} =\\
&e^{iNkx}\left(\frac{i\cos (2kx)}{2}\right)^{\frac{N}{2}}
\left(\frac{\cos (kx - \frac{\pi} {4})}{\cos (kx + \frac{\pi}{4})}
\right)^{N k x_{\mathrm{est}}}.
\label{eq:AnaBnb1}
\end{split}
\end{equation}
In the limit of small $k x$,
\begin{equation}
M_a(x)^{n_a } M_b(x)^{n_b}=\left(\frac{i}{2}\right)^{\frac{N}{2}}e^{N k^2
x_{\mathrm{est}}^2 } e^{iNkx} e^{ - Nk^2 \left(x -
x_{\mathrm{est}}\right)^2},
\label{eq:AnaBnb2}
\end{equation}
and the final wavefunction of the particle simplifies to
\begin{equation}
\psi_N(x) = \frac{e^{iNkx} e^{- Nk^2 \left(x - x_{\mathrm{est}}\right)^2}
\langle x |\psi_0\rangle}{\sqrt{\int\limits_{ - \infty }^\infty
{e^{- 2Nk^2 \left(x - x_{\mathrm{est}}\right)^2}|\langle x |\psi _0\rangle|^2 dx} }}.
\label{eq:ExplicitFinalWaveFunction}
\end{equation}
In the limit of large $N$, the post-measurement wavefunction is essentially a
Gaussian wavepacket centered about $x_{\rm{est}}$, and the corresponding
probability distribution for position has a variance of $1/(4Nk^2)=
\sigma^2_{x\rm{~est}}$, in precise agreement with the measured
position.  The wavefunction collapse is seen to be a direct consequence of
the cumulative outcome-specific backaction on the particle when entanglement
is destroyed by detection of photons.

\section{\label{sec:Averages}Probability density for $x_{\mathrm{est}}$}

Next, we find the probability of observing a particular data sequence in a
position measurement of a particle with initial wavefunction $\psi_0(x)$. The
probability for the sequence to occur is the product of the conditional
probabilities at each step:
\begin{equation}
 P_{\mathrm{seq}} = P(o_1 )P(o_2 |o_1 )...P(o_N |o_1 ,...,o_{N - 1} ).
\label{eq:SingleOutcomeProbability}
\end{equation}
This product is simplified by examining the $j^{th}$ factor.  From Eqs.\
(\ref{eq:ChannelProbabilities}) and (\ref{eq:FinalWaveFunction1}) we have
\begin{equation}
P(o_j|o_{j-1},\cdots)=\frac{\langle\psi_0|O_{1}^\dagger O_{2}^\dagger \cdots O_j^\dagger O_j \cdots O_{2}O_{1}|\psi _0\rangle}
{\langle\psi_0|O_{1}^\dagger O_{2}^\dagger \cdots O_{j-1}^\dagger O_{j-1} \cdots O_{2}O_{1}|\psi _0\rangle}.
\end{equation}
We see that the denominator of each factor in Eq.\ (\ref{eq:SingleOutcomeProbability})
cancels the numerator of the factor to its left, leaving
\begin{align}
P_{\mathrm{seq}} &= \langle\psi_0|(M_a^\dagger(x)M_a(x))^{n_a}(M_b^\dagger(x)M_b(x))^{n_b}|\psi_0\rangle\nonumber\\
&= \int\limits_{ - \infty }^\infty
{|M_a(x)^{n_a} M_b(x)^{n_b } \psi _0 (x)|^2 {\rm{d}}x}.
\label{eq:SingleOutcomeProbability2}
\end{align}

The probability of obtaining a particular data sequence depends on the number
of counts in each channel, but not the order in which the counts are
recorded. It follows that the probability of obtaining a particular value of
$n_b$ can be obtained by scaling Eq.\ (\ref{eq:SingleOutcomeProbability2})
with the appropriate binomial coefficient, which, applying Eq.\
(\ref{eq:AnaBnb2}), gives
\begin{equation}
P_{n_b} = \left(
\begin{array}{*{20}c}
   N  \\
   n_b\\
\end{array} \right)
\frac{e^{2 N k^2 x_{\mathrm{est}}^2}}{2^N } \int\limits_{ - \infty
}^\infty  e^{-2Nk^2(x-x_{\mathrm{est}})^2} \left| \psi_0 (x) \right|^2
{\rm{d}}x. \label{eq:EnsembleOutcomeProbability1}
\end{equation}
The right hand side of Eq. (\ref{eq:EnsembleOutcomeProbability1}) is
clarified by applying Sterling's approximation for the binomial
coefficients (see, for example, Ref.\ \onlinecite{Kittel80}) and expressing $n_b$ in terms of $x_{\rm{est}}$
and $N$, which leads to
\begin{equation}
P(x_{\rm{est}} ) = \int\limits_{ - \infty }^\infty
{\sqrt{\frac{2}{{\pi N}}} e^{-{\frac{{(x-x_{\rm{est}})^2}}
{{1/(2Nk^2)}}} }\left| {\psi_0 (x)}\right|^2 {\rm{d}}x.}
\label{eq:EnsembleOutcomeProbability3}
\end{equation}
For sufficiently large $N$, the initial wave function can be evaluated at
$x_{\rm{est}}$, leaving an integral that can be evaluated and giving us
\begin{equation}
P(x_{\rm{est}} ) = \frac{{\left| {\psi_0 (x_{\rm{est}} )} \right|^2}}
{{Nk}} = \left| {\psi_0 (x_{\rm{est}} )} \right|^2 \delta x_{\rm{est}}.
\label{eq:EnsembleOutcomProbability4}
\end{equation}
We find that the interpretation of the absolute square of the spatial wave
function as the probability density for the particle's position emerges from
the model.

\section{\label{sec:Momentum}Linear momentum of the post-measurement
wavefunction}
Next we consider the post-measurement expectation value of the particle's
linear momentum $\left\langle p_N\right\rangle$ and its variance
$\sigma_{p_N}^2$.  As $N$ increases, the influence of $\psi_0(x)$ becomes
relatively less important.  However, there is an absolute contribution that
persists, depending only on $\psi_0(x)$ and its derivatives evaluated at
$x = x_\mathrm{est}$. This persistent contribution conserves the particle's initial
linear momentum $\left\langle p_0\right\rangle$ and preserves its initial variance
$\sigma_{p_0}^2$ in the ensemble average.

In order to evaluate $\left\langle p_N\right\rangle$, we expand $\psi_0(x)$
as a second order power series in $(x-x_\mathrm{est})$:
\begin{equation}
\psi_0(x) \approx \left.\psi_{0}\right|_{x_{\rm{est}}} +
\left.\psi_{0}'\right|_{x_{\rm{est}}}(x-x_{\mathrm{est}}) +
\left.\frac{\psi_{0}''}{2}\right|_{x_{\rm{est}}}(x-x_{\mathrm{est}})^2.
\label{eq:powerserwf}
\end{equation}
Using Eq.\ (\ref{eq:powerserwf}) for $\psi_0(x)$ in Eq.\
(\ref{eq:ExplicitFinalWaveFunction}) gives
\begin{equation}
\left\langle p_N\right\rangle \approx \hbar k N -
\left.\frac{i\hbar\left(\psi_0^*\psi_0'-\psi _0{\psi_0'}^*\right)}
{2\left|{\psi_0}\right|^2}  \right|_{x_{\rm{est}} },
\label{eq:QuantumMomentum}
\end{equation}
where we have expanded the result as a series in powers of $1/N$ and dropped
terms that vanish in the absolute sense as $N$ increases. Expanding
$\psi_0(x)$ beyond second order does not affect the result.

The first term in Eq.\ (\ref{eq:QuantumMomentum}) arises from the elastic
collisions with half of the amplitude of the $N$ photons, each of which
imparts linear momentum $\hbar k$. The second term is the contribution to
$\left\langle p_0\right\rangle$ of $\psi_0(x)$ in the range $(x_\mathrm{est}
- \delta x_{\mathrm{est}}/2, x_\mathrm{est}+\delta x_{\mathrm{est}}/2)$.
Using the probability distribution (\ref{eq:EnsembleOutcomProbability4}), the
average over an ensemble of identical systems approaches
\begin{equation}
\left\langle p_N\right\rangle_{\mathrm{ens}} = \hbar k N + \left\langle
p_0\right\rangle.
\label{eq:EnsembleAverageMomentum}
\end{equation}

Eq.\ (\ref{eq:EnsembleAverageMomentum}) confirms conservation of linear
momentum in the measurement, but the conservation argument can be turned
around to motivate the de Broglie hypothesis for the wavelength of a
particle.  From Eq.\ (\ref{eq:ExplicitFinalWaveFunction}) $\psi_N(x)$ is a
wavepacket with wavelength
\begin{equation}
\lambda = \frac{2 \pi}{N k},
\label{eq:lambda}
\end{equation}
while conservation of linear momentum implies a post-measurement linear
momentum of
\begin{equation}
p=N \hbar k.
\label{eq:conp}
\end{equation}
Elimination of $Nk$ between Eqs. (\ref{eq:lambda}) and (\ref{eq:conp}) gives
the de Broglie result,
\begin{equation}
\lambda = \frac{h}{p},
\end{equation}
without explicit appeal to the calculation of $\left\langle p_N\right\rangle$
in the wave mechanics formalism.

A similar calculation gives $\sigma_{p_N}^2$ in the limit of large $N$.
Using Eq.\ (\ref{eq:powerserwf}) for $\psi_0(x)$ we find
\begin{equation}
\begin{split}
\sigma_{p_N}^2 \approx & N(\hbar k)^2 -
\left.\frac{\hbar^2\left(\psi _0^*\psi_0''+\psi_0{\psi_0''}^*-2{\psi _0'}^*
\psi_0'\right)}{4\left| \psi_0\right|^2}\right|_{x_{\mathrm{est}}} \\
& -\left(\left.\frac{-i\hbar\left(\psi_0^*\psi_0'-\psi _0{\psi _0'}^*
\right)}{2\left|\psi_0\right|^2}\right|_{x_{\mathrm{est}}}\right)^2.
\label{eq:varp}
\end{split}
\end{equation}
Note that keeping only the dominant term in Eq.\ (\ref{eq:varp}), and
referring to Eq.\ (\ref{eq:VarianceXEstimate}), the product
\begin{equation}
\sigma_{p_N}\sigma_{x_\mathrm{est}}=\frac{\hbar}{2}
\end{equation}
saturates Heisenberg's inequality for this measurement.

As $\sigma_{p_N}^2$ is not observable, the direct ensemble average of Eq.\
(\ref{eq:varp}) is not meaningful.  Taking care to evaluate the ensemble
average of the linear momentum before squaring, we find the post-measurement
ensemble variance of the particle's linear momentum approaches
\begin{equation}
\begin{split}
\sigma^2_{p_N\mathrm{ens}} &= N(\hbar k)^2\\
&+\left\langle\left.\frac{\hbar^2
\left(\psi _0^*\psi_0''+\psi_0{\psi_0''}^*-2{\psi_0'}^*\psi_0'\right)}
{4\left|\psi_0\right|^2}\right|_{x_{\mathrm{est}}}
\right\rangle_{\mathrm{ens}}\\
&-\left(\left\langle\left. \frac{i\hbar\left(
\psi_0^*\psi_0'-\psi _0{\psi _0'}^*\right)}
{2\left|{\psi_0}\right|^2}\right|_{x_{\mathrm{est}}}
\right\rangle_{\mathrm{ens}}\right)^2\\
&= N(\hbar k)^2+\left\langle p_0^2 \right\rangle -
\left\langle p_0 \right\rangle^2,
\label{eq:ensvarp}
\end{split}
\end{equation}
increasing in accordance with the uncertainty principle while preserving
$\sigma_{p_0}^2$ on average.

\section{\label{sec:WeakValues}Relationship to Weak Values}

Weak measurements are commonly associated with weak value amplification.
The weak measurements we describe above are not an example of weak value amplification in the normal sense.\cite{Aharonov1988apr, Ritchie1991mar, Dixon2009apr}
In particular, they do not include dark port tuning and the data are not limited to measurements on a postselected subset of the probe ancilla.
In contrast, our measurement procedure is optimized by maximizing the collection efficiency for all photon ancilla, and by tuning the photon detection probability for the two ports to be equal.
There is, however, a sense in which the detection of a photon at one or the other of the detectors could be characterized as the measurement of a weak value.
Referring to Eq.\ \ref{eq:xest} we could interpret the detection of a single photon in one of the detectors as a one-shot position measurement with $x_{\mathrm{est}} = \pm 1/2k$---much larger than what is allowed by the condition that $\psi_0(x)$ is localized in the region $|kx| << 1$.
Averaging of these amplified position values, of course, returns a realistic value for the position with high probability and the expected uncertainty in the mean.

\section{\label{sec:Conclusion}Discussion and Conclusion}
This work demonstrates a specific example of the decomposition of a continuous observable's projective measurement into a large number of weak interactions in the canonical context of position measurement by an optical interferometer.
By following the backaction of counting weakly interacting photons in detail, we reveal the connection between destructive detection of photons and the wave-like translational behavior of massive particles subject to ephemeral interactions.
The position probability density, momentum density, post-measurement wavefunction and de Broglie wavelength of the particle all agree with the predictions of elementary wave mechanics.
Interestingly, the relative weights of the terms involving $\left|\psi_0'\right|^2$ and $\mathrm{Re}(\psi_0^*\psi_0'')$ in Eq.\ (\ref{eq:ensvarp}) that together average to $\langle p_0^2 \rangle$ were not readily anticipated and merit further study.
These results demonstrate that local mechanisms such as environmental decoherence, spontaneous local collapse or hidden observers acting directly on the particle are not strictly necessary for spacial wavefunction collapse of a remote particle to occur.

Our semiclassical treatment of the optical modes in this analysis highlights the essential elements of a quantum mechanical position measurement.  By adopting this approach, we treat photon counting as a primitive operation and thereby cut off the otherwise infinite regression of treating a stochastic measurement process on one state space as increasingly detailed unitary evolutions on ever larger state spaces.  Such a cut off is necessary in any quantum analysis that purports to describe recorded data, whether it is done implicitly or explicitly.  This semiclassical approach also provides a starting point for obtaining quantitative predictions of, for example, the power spectra of a continuously monitored position signal and the resulting backaction force through more sophisticated analyses that account for the evolution of the particle state during the necessarily finite measurement time, the finite photon line width, the detector performance, and the full second quantization of the optical field.\cite{Gough2015jan, Gough2016jan}
Finally, from the reasoning that leads to the inclusion of the binomial coefficient in Eq. (28), we note that our model demonstrates how a particular outcome for the measured value of the position and the post-measurement quantum state of a particle can result from numerous distinct trajectories through state space.
When more complex interferometer-based position measurements that allow for temporal delays in photon detection through, e.g., spatial separation of the detectors are considered, it becomes possible that observers in relative motion will disagree on the order in which photon counts are registered and hence the particle's state space trajectory.
The need to accommodate observer-dependent state space trajectories, even in the case of pure states, may provide insight into why the quantum evolution is not fully explained by a single law of motion.
%Finally, we note that models such as the one developed here demonstrate how various trajectories through state space can have the same outcome and probability and that the study of such models, including the investigation of different tunings, post measurement tunings, delays in photon detection, and spatial separation of the detectors from the particle and each other, could further our understanding of how a physical system can accommodate a fundamental theory in which the evolution of the state is governed simultaneously by two apparently incompatible laws of motion.

\begin{acknowledgments}
I would like to thank John Sidles and Joseph Garbini of the University of Washington for for inspiring this work during visits to 
%starting me on the path that lead to this work while visiting
their labs during July 2005 and June 2012 by teaching me how to develop the measurement operators for the passage of a single photon through an optical interferometer and encouraging me to explore the implications of performing such a measurement iteratively.
I would also like to thank the organizers and participants of the international workshop, “Magnetic Resonance Force Microscopy: Routes to Three-Dimensional Imaging of Single Molecules,” held on June 21-24, 2006, at Cornell University, for many helpful discussions.
The workshop was supported by grants from the Kavli Institute at Cornell for Nanoscale Science and the National Science Foundation Under Grant No.\ DMR-0634455, with additional funding from IBM, Eastman Kodak, and the New York State Office of Science, Technology, and Academic Research.
Additional support for this work was provided by the Department of Physics and Nuclear Engineering and the Mathematical Sciences Center of Excellence of the U.\ S.\ Military Academy.
\end{acknowledgments}

%%%%%%%%%%%%%%%%%%%%%%%%%%%%%%%%%%%%%%%%%%%%%%%%%%%%%%%%%%%%%%%%%
\bibliographystyle{apsrev4-1}
\bibliography{collapse}

%merlin.mbs apsrev4-1.bst 2010-07-25 4.21a (PWD, AO, DPC) hacked
%Control: key (0)
%Control: author (72) initials jnrlst
%Control: editor formatted (1) identically to author
%Control: production of article title (-1) disabled
%Control: page (0) single
%Control: year (1) truncated
%Control: production of eprint (0) enabled
\begin{thebibliography}{40}%
\makeatletter
\providecommand \@ifxundefined [1]{%
 \@ifx{#1\undefined}
}%
\providecommand \@ifnum [1]{%
 \ifnum #1\expandafter \@firstoftwo
 \else \expandafter \@secondoftwo
 \fi
}%
\providecommand \@ifx [1]{%
 \ifx #1\expandafter \@firstoftwo
 \else \expandafter \@secondoftwo
 \fi
}%
\providecommand \natexlab [1]{#1}%
\providecommand \enquote  [1]{``#1''}%
\providecommand \bibnamefont  [1]{#1}%
\providecommand \bibfnamefont [1]{#1}%
\providecommand \citenamefont [1]{#1}%
\providecommand \href@noop [0]{\@secondoftwo}%
\providecommand \href [0]{\begingroup \@sanitize@url \@href}%
\providecommand \@href[1]{\@@startlink{#1}\@@href}%
\providecommand \@@href[1]{\endgroup#1\@@endlink}%
\providecommand \@sanitize@url [0]{\catcode `\\12\catcode `\$12\catcode
  `\&12\catcode `\#12\catcode `\^12\catcode `\_12\catcode `\%12\relax}%
\providecommand \@@startlink[1]{}%
\providecommand \@@endlink[0]{}%
\providecommand \url  [0]{\begingroup\@sanitize@url \@url }%
\providecommand \@url [1]{\endgroup\@href {#1}{\urlprefix }}%
\providecommand \urlprefix  [0]{URL }%
\providecommand \Eprint [0]{\href }%
\providecommand \doibase [0]{http://dx.doi.org/}%
\providecommand \selectlanguage [0]{\@gobble}%
\providecommand \bibinfo  [0]{\@secondoftwo}%
\providecommand \bibfield  [0]{\@secondoftwo}%
\providecommand \translation [1]{[#1]}%
\providecommand \BibitemOpen [0]{}%
\providecommand \bibitemStop [0]{}%
\providecommand \bibitemNoStop [0]{.\EOS\space}%
\providecommand \EOS [0]{\spacefactor3000\relax}%
\providecommand \BibitemShut  [1]{\csname bibitem#1\endcsname}%
\let\auto@bib@innerbib\@empty
%</preamble>
\bibitem [{\citenamefont {L\"uders}(2006)}]{Luders2006sep}%
  \BibitemOpen
  \bibfield  {author} {\bibinfo {author} {\bibfnamefont {G.}~\bibnamefont
  {L\"uders}},\ }\href {\doibase 10.1002/andp.200610207} {\bibfield  {journal}
  {\bibinfo  {journal} {Ann. Phys.}\ }\textbf {\bibinfo {volume} {15}},\
  \bibinfo {pages} {663 } (\bibinfo {year} {2006})}\BibitemShut {NoStop}%
\bibitem [{\citenamefont {Ru\v{z}a}(2010)}]{Ruza2010jan}%
  \BibitemOpen
  \bibfield  {author} {\bibinfo {author} {\bibfnamefont {J.}~\bibnamefont
  {Ru\v{z}a}},\ }\href {\doibase 10.1016/j.physe.2009.06.052} {\bibfield
  {journal} {\bibinfo  {journal} {Physica E: Low-dimensional Systems and
  Nanostructures}\ }\textbf {\bibinfo {volume} {42}},\ \bibinfo {pages} {330 }
  (\bibinfo {year} {2010})}\BibitemShut {NoStop}%
\bibitem [{\citenamefont {Allahverdyan}\ \emph {et~al.}(2013)\citenamefont
  {Allahverdyan}, \citenamefont {Balian},\ and\ \citenamefont
  {Nieuwenhuizen}}]{Allahverdyan2013apr}%
  \BibitemOpen
  \bibfield  {author} {\bibinfo {author} {\bibfnamefont {A.~E.}\ \bibnamefont
  {Allahverdyan}}, \bibinfo {author} {\bibfnamefont {R.}~\bibnamefont
  {Balian}}, \ and\ \bibinfo {author} {\bibfnamefont {T.~M.}\ \bibnamefont
  {Nieuwenhuizen}},\ }\href {\doibase 10.1016/j.physrep.2012.11.001} {\bibfield
   {journal} {\bibinfo  {journal} {Physics Reports}\ }\textbf {\bibinfo
  {volume} {525}},\ \bibinfo {pages} {1 } (\bibinfo {year} {2013})}\BibitemShut
  {NoStop}%
\bibitem [{\citenamefont {Bassi}\ and\ \citenamefont
  {Ghirardi}(2003)}]{Bassi03}%
  \BibitemOpen
  \bibfield  {author} {\bibinfo {author} {\bibfnamefont {A.}~\bibnamefont
  {Bassi}}\ and\ \bibinfo {author} {\bibfnamefont {G.}~\bibnamefont
  {Ghirardi}},\ }\href {\doibase 10.1016/S0370-1573(03)00103-0} {\bibfield
  {journal} {\bibinfo  {journal} {Physics Reports}\ }\textbf {\bibinfo {volume}
  {379}},\ \bibinfo {pages} {257 } (\bibinfo {year} {2003})}\BibitemShut
  {NoStop}%
\bibitem [{\citenamefont {Schlosshauer}(2005)}]{Schlosshauer04}%
  \BibitemOpen
  \bibfield  {author} {\bibinfo {author} {\bibfnamefont {M.}~\bibnamefont
  {Schlosshauer}},\ }\href {\doibase 10.1103/RevModPhys.76.1267} {\bibfield
  {journal} {\bibinfo  {journal} {Rev. Mod. Phys.}\ }\textbf {\bibinfo {volume}
  {76}},\ \bibinfo {pages} {1267 } (\bibinfo {year} {2005})}\BibitemShut
  {NoStop}%
\bibitem [{\citenamefont {Weinberg}(2012)}]{Weinberg2012jun}%
  \BibitemOpen
  \bibfield  {author} {\bibinfo {author} {\bibfnamefont {S.}~\bibnamefont
  {Weinberg}},\ }\href {\doibase 10.1103/PhysRevA.85.062116} {\bibfield
  {journal} {\bibinfo  {journal} {Phys. Rev. A}\ }\textbf {\bibinfo {volume}
  {85}},\ \bibinfo {pages} {062116} (\bibinfo {year} {2012})}\BibitemShut
  {NoStop}%
\bibitem [{\citenamefont {Bassi}\ \emph {et~al.}(2013)\citenamefont {Bassi},
  \citenamefont {Lochan}, \citenamefont {Satin}, \citenamefont {Singh},\ and\
  \citenamefont {Ulbricht}}]{Bassi13}%
  \BibitemOpen
  \bibfield  {author} {\bibinfo {author} {\bibfnamefont {A.}~\bibnamefont
  {Bassi}}, \bibinfo {author} {\bibfnamefont {K.}~\bibnamefont {Lochan}},
  \bibinfo {author} {\bibfnamefont {S.}~\bibnamefont {Satin}}, \bibinfo
  {author} {\bibfnamefont {T.~P.}\ \bibnamefont {Singh}}, \ and\ \bibinfo
  {author} {\bibfnamefont {H.}~\bibnamefont {Ulbricht}},\ }\href {\doibase
  10.1103/RevModPhys.85.471} {\bibfield  {journal} {\bibinfo  {journal} {Rev.
  Mod. Phys.}\ }\textbf {\bibinfo {volume} {85}},\ \bibinfo {pages} {471 }
  (\bibinfo {year} {2013})}\BibitemShut {NoStop}%
\bibitem [{\citenamefont {David~Mermin}(1989)}]{mermin89}%
  \BibitemOpen
  \bibfield  {author} {\bibinfo {author} {\bibfnamefont {N.}~\bibnamefont
  {David~Mermin}},\ }\href {\doibase 10.1063/1.2810963} {\bibfield  {journal}
  {\bibinfo  {journal} {Phys. Today}\ }\textbf {\bibinfo {volume} {42}},\
  \bibinfo {pages} {9} (\bibinfo {year} {1989})}\BibitemShut {NoStop}%
\bibitem [{\citenamefont {Nimmrichter}\ \emph {et~al.}(2014)\citenamefont
  {Nimmrichter}, \citenamefont {Hornberger},\ and\ \citenamefont
  {Hammerer}}]{Nimmrichter2014jul}%
  \BibitemOpen
  \bibfield  {author} {\bibinfo {author} {\bibfnamefont {S.}~\bibnamefont
  {Nimmrichter}}, \bibinfo {author} {\bibfnamefont {K.}~\bibnamefont
  {Hornberger}}, \ and\ \bibinfo {author} {\bibfnamefont {K.}~\bibnamefont
  {Hammerer}},\ }\href {\doibase 10.1103/PhysRevLett.113.020405} {\bibfield
  {journal} {\bibinfo  {journal} {Phys. Rev. Lett.}\ }\textbf {\bibinfo
  {volume} {113}},\ \bibinfo {pages} {020405} (\bibinfo {year}
  {2014})}\BibitemShut {NoStop}%
\bibitem [{\citenamefont {Di\'osi}(2015)}]{Diosi2015feb}%
  \BibitemOpen
  \bibfield  {author} {\bibinfo {author} {\bibfnamefont {L.}~\bibnamefont
  {Di\'osi}},\ }\href {\doibase 10.1103/PhysRevLett.114.050403} {\bibfield
  {journal} {\bibinfo  {journal} {Phys. Rev. Lett.}\ }\textbf {\bibinfo
  {volume} {114}},\ \bibinfo {pages} {050403} (\bibinfo {year}
  {2015})}\BibitemShut {NoStop}%
\bibitem [{\citenamefont {Hohenberg}(2010)}]{Hohenberg2010oct}%
  \BibitemOpen
  \bibfield  {author} {\bibinfo {author} {\bibfnamefont {P.~C.}\ \bibnamefont
  {Hohenberg}},\ }\href {\doibase 10.1103/RevModPhys.82.2835} {\bibfield
  {journal} {\bibinfo  {journal} {Rev. Mod. Phys.}\ }\textbf {\bibinfo {volume}
  {82}},\ \bibinfo {pages} {2835 } (\bibinfo {year} {2010})}\BibitemShut
  {NoStop}%
\bibitem [{\citenamefont {Fuchs}\ and\ \citenamefont {Schack}(2013)}]{Fuchs13}%
  \BibitemOpen
  \bibfield  {author} {\bibinfo {author} {\bibfnamefont {C.~A.}\ \bibnamefont
  {Fuchs}}\ and\ \bibinfo {author} {\bibfnamefont {R.}~\bibnamefont {Schack}},\
  }\href {\doibase 10.1103/RevModPhys.85.1693} {\bibfield  {journal} {\bibinfo
  {journal} {Rev. Mod. Phys.}\ }\textbf {\bibinfo {volume} {85}},\ \bibinfo
  {pages} {1693 } (\bibinfo {year} {2013})}\BibitemShut {NoStop}%
\bibitem [{\citenamefont {Zurek}(2003)}]{Zurek2003may}%
  \BibitemOpen
  \bibfield  {author} {\bibinfo {author} {\bibfnamefont {W.~H.}\ \bibnamefont
  {Zurek}},\ }\href {\doibase 10.1103/RevModPhys.75.715} {\bibfield  {journal}
  {\bibinfo  {journal} {Rev. Mod. Phys.}\ }\textbf {\bibinfo {volume} {75}},\
  \bibinfo {pages} {715 } (\bibinfo {year} {2003})}\BibitemShut {NoStop}%
\bibitem [{\citenamefont {Pernice}\ and\ \citenamefont
  {Strunz}(2011)}]{Pernice2011dec}%
  \BibitemOpen
  \bibfield  {author} {\bibinfo {author} {\bibfnamefont {A.}~\bibnamefont
  {Pernice}}\ and\ \bibinfo {author} {\bibfnamefont {W.~T.}\ \bibnamefont
  {Strunz}},\ }\href {\doibase 10.1103/PhysRevA.84.062121} {\bibfield
  {journal} {\bibinfo  {journal} {Phys. Rev. A}\ }\textbf {\bibinfo {volume}
  {84}},\ \bibinfo {pages} {062121} (\bibinfo {year} {2011})}\BibitemShut
  {NoStop}%
\bibitem [{\citenamefont {Parisio}(2011)}]{Parisio2011dec}%
  \BibitemOpen
  \bibfield  {author} {\bibinfo {author} {\bibfnamefont {F.}~\bibnamefont
  {Parisio}},\ }\href {\doibase 10.1103/PhysRevA.84.062108} {\bibfield
  {journal} {\bibinfo  {journal} {Phys. Rev. A}\ }\textbf {\bibinfo {volume}
  {84}},\ \bibinfo {pages} {062108} (\bibinfo {year} {2011})}\BibitemShut
  {NoStop}%
\bibitem [{\citenamefont {Fleming}(2013)}]{Fleming13}%
  \BibitemOpen
  \bibfield  {author} {\bibinfo {author} {\bibfnamefont {A.~J.}\ \bibnamefont
  {Fleming}},\ }\href {\doibase 10.1016/j.sna.2012.10.016} {\bibfield
  {journal} {\bibinfo  {journal} {Sensors and Actuators A: Physical}\ }\textbf
  {\bibinfo {volume} {190}},\ \bibinfo {pages} {106 } (\bibinfo {year}
  {2013})}\BibitemShut {NoStop}%
\bibitem [{\citenamefont {Loudon}(2003)}]{Loudon03}%
  \BibitemOpen
  \bibfield  {author} {\bibinfo {author} {\bibfnamefont {R.}~\bibnamefont
  {Loudon}},\ }\href@noop {} {\emph {\bibinfo {title} {The Quantum Theory of
  Light, 3rd Ed. (Digital)}}}\ (\bibinfo  {publisher} {Oxford University
  Press},\ \bibinfo {address} {Oxford},\ \bibinfo {year} {2003})\BibitemShut
  {NoStop}%
\bibitem [{\citenamefont {von Neumann}(1932)}]{vonN32}%
  \BibitemOpen
  \bibfield  {author} {\bibinfo {author} {\bibfnamefont {J.}~\bibnamefont {von
  Neumann}},\ }\href@noop {} {\emph {\bibinfo {title} {Mathematische Grundlagen
  der Quantenmechanil}}}\ (\bibinfo  {publisher} {Springer},\ \bibinfo
  {address} {Berlin},\ \bibinfo {year} {1932})\BibitemShut {NoStop}%
\bibitem [{\citenamefont {Nielsen}\ and\ \citenamefont
  {Chuang}(2000)}]{Nielsen00}%
  \BibitemOpen
  \bibfield  {author} {\bibinfo {author} {\bibfnamefont {M.~A.}\ \bibnamefont
  {Nielsen}}\ and\ \bibinfo {author} {\bibfnamefont {I.~L.}\ \bibnamefont
  {Chuang}},\ }\href@noop {} {\emph {\bibinfo {title} {Quantum Computation and
  Quantum Information}}}\ (\bibinfo  {publisher} {Cambridge University Press},\
  \bibinfo {address} {Cambridge},\ \bibinfo {year} {2000})\BibitemShut
  {NoStop}%
\bibitem [{\citenamefont {Wiseman}\ and\ \citenamefont
  {Milburn}(2009)}]{Wiseman09}%
  \BibitemOpen
  \bibfield  {author} {\bibinfo {author} {\bibfnamefont {H.~M.}\ \bibnamefont
  {Wiseman}}\ and\ \bibinfo {author} {\bibfnamefont {G.~J.}\ \bibnamefont
  {Milburn}},\ }\href@noop {} {\emph {\bibinfo {title} {Quantum Measurement and
  Coltrol}}}\ (\bibinfo  {publisher} {Cambridge University Press},\ \bibinfo
  {address} {Cambridge},\ \bibinfo {year} {2009})\BibitemShut {NoStop}%
\bibitem [{\citenamefont {Clerk}\ \emph {et~al.}(2010)\citenamefont {Clerk},
  \citenamefont {Devoret}, \citenamefont {Girvin}, \citenamefont {Marquardt},\
  and\ \citenamefont {Schoelkopf}}]{Clerk2010apr}%
  \BibitemOpen
  \bibfield  {author} {\bibinfo {author} {\bibfnamefont {A.~A.}\ \bibnamefont
  {Clerk}}, \bibinfo {author} {\bibfnamefont {M.~H.}\ \bibnamefont {Devoret}},
  \bibinfo {author} {\bibfnamefont {S.~M.}\ \bibnamefont {Girvin}}, \bibinfo
  {author} {\bibfnamefont {F.}~\bibnamefont {Marquardt}}, \ and\ \bibinfo
  {author} {\bibfnamefont {R.~J.}\ \bibnamefont {Schoelkopf}},\ }\href
  {\doibase 10.1103/RevModPhys.82.1155} {\bibfield  {journal} {\bibinfo
  {journal} {Rev. Mod. Phys.}\ }\textbf {\bibinfo {volume} {82}},\ \bibinfo
  {pages} {1155 } (\bibinfo {year} {2010})}\BibitemShut {NoStop}%
\bibitem [{\citenamefont {Jacobs}(2014)}]{Jacobs14}%
  \BibitemOpen
  \bibfield  {author} {\bibinfo {author} {\bibfnamefont {K.}~\bibnamefont
  {Jacobs}},\ }\href@noop {} {\emph {\bibinfo {title} {Quantum Measurement
  Theory and its Applications}}}\ (\bibinfo  {publisher} {Cambridge University
  Press},\ \bibinfo {address} {Cambridge},\ \bibinfo {year} {2014})\BibitemShut
  {NoStop}%
\bibitem [{\citenamefont {Oreshkov}\ and\ \citenamefont
  {Brun}(2005)}]{Oreshkov05}%
  \BibitemOpen
  \bibfield  {author} {\bibinfo {author} {\bibfnamefont {O.}~\bibnamefont
  {Oreshkov}}\ and\ \bibinfo {author} {\bibfnamefont {T.~A.}\ \bibnamefont
  {Brun}},\ }\href {\doibase 10.1103/PhysRevLett.95.110409} {\bibfield
  {journal} {\bibinfo  {journal} {Phys. Rev. Lett.}\ }\textbf {\bibinfo
  {volume} {95}},\ \bibinfo {pages} {110409} (\bibinfo {year}
  {2005})}\BibitemShut {NoStop}%
\bibitem [{\citenamefont {Varbanov}\ and\ \citenamefont
  {Brun}(2007)}]{Varbanov2007sep}%
  \BibitemOpen
  \bibfield  {author} {\bibinfo {author} {\bibfnamefont {M.}~\bibnamefont
  {Varbanov}}\ and\ \bibinfo {author} {\bibfnamefont {T.~A.}\ \bibnamefont
  {Brun}},\ }\href {\doibase 10.1103/PhysRevA.76.032104} {\bibfield  {journal}
  {\bibinfo  {journal} {Phys. Rev. A}\ }\textbf {\bibinfo {volume} {76}},\
  \bibinfo {pages} {032104} (\bibinfo {year} {2007})}\BibitemShut {NoStop}%
\bibitem [{\citenamefont {Gisin}(1984)}]{Gisin1984may}%
  \BibitemOpen
  \bibfield  {author} {\bibinfo {author} {\bibfnamefont {N.}~\bibnamefont
  {Gisin}},\ }\href {\doibase 10.1103/PhysRevLett.52.1657} {\bibfield
  {journal} {\bibinfo  {journal} {Phys. Rev. Lett.}\ }\textbf {\bibinfo
  {volume} {52}},\ \bibinfo {pages} {1657 } (\bibinfo {year}
  {1984})}\BibitemShut {NoStop}%
\bibitem [{\citenamefont {Gisin}(1989)}]{Gisin1989}%
  \BibitemOpen
  \bibfield  {author} {\bibinfo {author} {\bibfnamefont {N.}~\bibnamefont
  {Gisin}},\ }\href@noop {} {\bibfield  {journal} {\bibinfo  {journal}
  {Helvetica Physica Acta}\ }\textbf {\bibinfo {volume} {62}},\ \bibinfo
  {pages} {363} (\bibinfo {year} {1989})}\BibitemShut {NoStop}%
\bibitem [{\citenamefont {Belavkin}(1992)}]{Belavkin1992jun}%
  \BibitemOpen
  \bibfield  {author} {\bibinfo {author} {\bibfnamefont {V.~P.}\ \bibnamefont
  {Belavkin}},\ }\href {\doibase 10.1007/BF02097018} {\bibfield  {journal}
  {\bibinfo  {journal} {Commun.Math. Phys.}\ }\textbf {\bibinfo {volume}
  {146}},\ \bibinfo {pages} {611 } (\bibinfo {year} {1992})}\BibitemShut
  {NoStop}%
\bibitem [{\citenamefont {Brun}(2002)}]{Brun2002unk}%
  \BibitemOpen
  \bibfield  {author} {\bibinfo {author} {\bibfnamefont {T.~A.}\ \bibnamefont
  {Brun}},\ }\href {\doibase 10.1119/1.1475328} {\bibfield  {journal} {\bibinfo
   {journal} {Am. J. Phys.}\ }\textbf {\bibinfo {volume} {70}},\ \bibinfo
  {pages} {719} (\bibinfo {year} {2002})}\BibitemShut {NoStop}%
\bibitem [{\citenamefont {Cheong}\ and\ \citenamefont
  {Lee}(2012)}]{Cheong2012oct}%
  \BibitemOpen
  \bibfield  {author} {\bibinfo {author} {\bibfnamefont {Y.~W.}\ \bibnamefont
  {Cheong}}\ and\ \bibinfo {author} {\bibfnamefont {S.-W.}\ \bibnamefont
  {Lee}},\ }\href {\doibase 10.1103/PhysRevLett.109.150402} {\bibfield
  {journal} {\bibinfo  {journal} {Phys. Rev. Lett.}\ }\textbf {\bibinfo
  {volume} {109}},\ \bibinfo {pages} {150402} (\bibinfo {year}
  {2012})}\BibitemShut {NoStop}%
\bibitem [{\citenamefont {Aharonov}\ \emph {et~al.}(1988)\citenamefont
  {Aharonov}, \citenamefont {Albert},\ and\ \citenamefont
  {Vaidman}}]{Aharonov1988apr}%
  \BibitemOpen
  \bibfield  {author} {\bibinfo {author} {\bibfnamefont {Y.}~\bibnamefont
  {Aharonov}}, \bibinfo {author} {\bibfnamefont {D.~Z.}\ \bibnamefont
  {Albert}}, \ and\ \bibinfo {author} {\bibfnamefont {L.}~\bibnamefont
  {Vaidman}},\ }\href {\doibase 10.1103/PhysRevLett.60.1351} {\bibfield
  {journal} {\bibinfo  {journal} {Phys. Rev. Lett.}\ }\textbf {\bibinfo
  {volume} {60}},\ \bibinfo {pages} {1351 } (\bibinfo {year}
  {1988})}\BibitemShut {NoStop}%
\bibitem [{\citenamefont {Ritchie}\ \emph {et~al.}(1991)\citenamefont
  {Ritchie}, \citenamefont {Story},\ and\ \citenamefont
  {Hulet}}]{Ritchie1991mar}%
  \BibitemOpen
  \bibfield  {author} {\bibinfo {author} {\bibfnamefont {N.~W.~M.}\
  \bibnamefont {Ritchie}}, \bibinfo {author} {\bibfnamefont {J.~G.}\
  \bibnamefont {Story}}, \ and\ \bibinfo {author} {\bibfnamefont {R.~G.}\
  \bibnamefont {Hulet}},\ }\href {\doibase 10.1103/PhysRevLett.66.1107}
  {\bibfield  {journal} {\bibinfo  {journal} {Phys. Rev. Lett.}\ }\textbf
  {\bibinfo {volume} {66}},\ \bibinfo {pages} {1107 } (\bibinfo {year}
  {1991})}\BibitemShut {NoStop}%
\bibitem [{\citenamefont {Dixon}\ \emph {et~al.}(2009)\citenamefont {Dixon},
  \citenamefont {Starling}, \citenamefont {Jordan},\ and\ \citenamefont
  {Howell}}]{Dixon2009apr}%
  \BibitemOpen
  \bibfield  {author} {\bibinfo {author} {\bibfnamefont {P.~B.}\ \bibnamefont
  {Dixon}}, \bibinfo {author} {\bibfnamefont {D.~J.}\ \bibnamefont {Starling}},
  \bibinfo {author} {\bibfnamefont {A.~N.}\ \bibnamefont {Jordan}}, \ and\
  \bibinfo {author} {\bibfnamefont {J.~C.}\ \bibnamefont {Howell}},\ }\href
  {\doibase 10.1103/PhysRevLett.102.173601} {\bibfield  {journal} {\bibinfo
  {journal} {Phys. Rev. Lett.}\ }\textbf {\bibinfo {volume} {102}},\ \bibinfo
  {pages} {173601} (\bibinfo {year} {2009})}\BibitemShut {NoStop}%
\bibitem [{\citenamefont {Starling}\ \emph {et~al.}(2009)\citenamefont
  {Starling}, \citenamefont {Dixon}, \citenamefont {Jordan},\ and\
  \citenamefont {Howell}}]{Starling2009oct}%
  \BibitemOpen
  \bibfield  {author} {\bibinfo {author} {\bibfnamefont {D.~J.}\ \bibnamefont
  {Starling}}, \bibinfo {author} {\bibfnamefont {P.~B.}\ \bibnamefont {Dixon}},
  \bibinfo {author} {\bibfnamefont {A.~N.}\ \bibnamefont {Jordan}}, \ and\
  \bibinfo {author} {\bibfnamefont {J.~C.}\ \bibnamefont {Howell}},\ }\href
  {\doibase 10.1103/PhysRevA.80.041803} {\bibfield  {journal} {\bibinfo
  {journal} {Phys. Rev. A}\ }\textbf {\bibinfo {volume} {80}},\ \bibinfo
  {pages} {041803} (\bibinfo {year} {2009})}\BibitemShut {NoStop}%
\bibitem [{\citenamefont {Nishizawa}\ \emph {et~al.}(2012)\citenamefont
  {Nishizawa}, \citenamefont {Nakamura},\ and\ \citenamefont
  {Fujimoto}}]{Nishizawa2012jun}%
  \BibitemOpen
  \bibfield  {author} {\bibinfo {author} {\bibfnamefont {A.}~\bibnamefont
  {Nishizawa}}, \bibinfo {author} {\bibfnamefont {K.}~\bibnamefont {Nakamura}},
  \ and\ \bibinfo {author} {\bibfnamefont {M.-K.}\ \bibnamefont {Fujimoto}},\
  }\href {\doibase 10.1103/PhysRevA.85.062108} {\bibfield  {journal} {\bibinfo
  {journal} {Phys. Rev. A}\ }\textbf {\bibinfo {volume} {85}},\ \bibinfo
  {pages} {062108} (\bibinfo {year} {2012})}\BibitemShut {NoStop}%
\bibitem [{\citenamefont {Rozema}\ \emph {et~al.}(2012)\citenamefont {Rozema},
  \citenamefont {Darabi}, \citenamefont {Mahler}, \citenamefont {Hayat},
  \citenamefont {Soudagar},\ and\ \citenamefont {Steinberg}}]{Rozema2012sep}%
  \BibitemOpen
  \bibfield  {author} {\bibinfo {author} {\bibfnamefont {L.~A.}\ \bibnamefont
  {Rozema}}, \bibinfo {author} {\bibfnamefont {A.}~\bibnamefont {Darabi}},
  \bibinfo {author} {\bibfnamefont {D.~H.}\ \bibnamefont {Mahler}}, \bibinfo
  {author} {\bibfnamefont {A.}~\bibnamefont {Hayat}}, \bibinfo {author}
  {\bibfnamefont {Y.}~\bibnamefont {Soudagar}}, \ and\ \bibinfo {author}
  {\bibfnamefont {A.~M.}\ \bibnamefont {Steinberg}},\ }\href {\doibase
  10.1103/PhysRevLett.109.100404} {\bibfield  {journal} {\bibinfo  {journal}
  {Phys. Rev. Lett.}\ }\textbf {\bibinfo {volume} {109}},\ \bibinfo {pages}
  {100404} (\bibinfo {year} {2012})}\BibitemShut {NoStop}%
\bibitem [{\citenamefont {Lund}\ and\ \citenamefont
  {Wiseman}(2010)}]{Lund2010sep}%
  \BibitemOpen
  \bibfield  {author} {\bibinfo {author} {\bibfnamefont {A.~P.}\ \bibnamefont
  {Lund}}\ and\ \bibinfo {author} {\bibfnamefont {H.~M.}\ \bibnamefont
  {Wiseman}},\ }\href {\doibase 10.1088/1367-2630/12/9/093011} {\bibfield
  {journal} {\bibinfo  {journal} {New J. Phys.}\ }\textbf {\bibinfo {volume}
  {12}},\ \bibinfo {pages} {093011} (\bibinfo {year} {2010})}\BibitemShut
  {NoStop}%
\bibitem [{\citenamefont {D'Ariano}\ and\ \citenamefont
  {Sacchi}(1997)}]{DAriano1997jul}%
  \BibitemOpen
  \bibfield  {author} {\bibinfo {author} {\bibfnamefont {G.}~\bibnamefont
  {D'Ariano}}\ and\ \bibinfo {author} {\bibfnamefont {M.}~\bibnamefont
  {Sacchi}},\ }\href {\doibase 10.1016/S0375-9601(97)00314-9} {\bibfield
  {journal} {\bibinfo  {journal} {Physics Letters A}\ }\textbf {\bibinfo
  {volume} {231}},\ \bibinfo {pages} {325 } (\bibinfo {year}
  {1997})}\BibitemShut {NoStop}%
\bibitem [{\citenamefont {Kittel}\ and\ \citenamefont
  {Kroemer}(1980)}]{Kittel80}%
  \BibitemOpen
  \bibfield  {author} {\bibinfo {author} {\bibfnamefont {C.}~\bibnamefont
  {Kittel}}\ and\ \bibinfo {author} {\bibfnamefont {H.}~\bibnamefont
  {Kroemer}},\ }\href@noop {} {\emph {\bibinfo {title} {Thermal Physics, 2nd
  Ed.}}}\ (\bibinfo  {publisher} {W. H. Freeman and Company},\ \bibinfo
  {address} {New York},\ \bibinfo {year} {1980})\ \bibinfo {note} {pp.
  10-23}\BibitemShut {NoStop}%
\bibitem [{\citenamefont {Gough}(2015)}]{Gough2015jan}%
  \BibitemOpen
  \bibfield  {author} {\bibinfo {author} {\bibfnamefont {J.~E.}\ \bibnamefont
  {Gough}},\ }\href {\doibase 10.1103/PhysRevA.91.013802} {\bibfield  {journal}
  {\bibinfo  {journal} {Phys. Rev. A}\ }\textbf {\bibinfo {volume} {91}},\
  \bibinfo {pages} {013802} (\bibinfo {year} {2015})}\BibitemShut {NoStop}%
\bibitem [{\citenamefont {Gough}(2016)}]{Gough2016jan}%
  \BibitemOpen
  \bibfield  {author} {\bibinfo {author} {\bibfnamefont {J.~E.}\ \bibnamefont
  {Gough}},\ }\href@noop {} {\enquote {\bibinfo {title} {Interferometric phase
  estimation though quantum filtering in coherent states},}\ } (\bibinfo {year}
  {2016}),\ \bibinfo {note} {arXiv:1601.04374[quant-ph]}\BibitemShut {NoStop}%
\end{thebibliography}%

\end{document}